\documentclass[prd,amssymb,twocolumn,10pt]{revtex4}
\usepackage{amsmath,amsfonts,amssymb}
\usepackage{verbatim,graphicx,float}

\newcommand{\dR}{\mathbb R}

\begin{document}

\title{Observables for FRW model with cosmological constant \\in the framework of loop cosmology}

\author{Jakub Mielczarek}
\email{jakub.mielczarek@uj.edu.pl}
\affiliation{Astronomical Observatory, Jagiellonian University, 30-244
Krak\'ow, Orla 171, Poland}

\author{W{\l}odzimierz Piechocki}
\email{piech@fuw.edu.pl}
\affiliation{Theoretical Physics Department, Institute for Nuclear Studies,
00-681 Warszawa, Ho{\.z}a 69, Poland}

\date{\today}

\begin{abstract}
We consider a flat cosmological model with a free massless scalar
field and the cosmological constant  $\Lambda$  in the framework
of loop quantum cosmology. The scalar field plays the role of an
intrinsic time. We apply the reduced phase space approach. The
dynamics of the model is solved analytically. We identify
elementary observables and their algebra. The compound physical
observables like the volume and the energy density of matter field
are analysed.  Both compound observables are bounded and oscillate
in the $\Lambda<0$  case. The energy density is bounded and
oscillates in the $\Lambda>0$  case. However, the volume is
unbounded from above, but periodic.  The difference between
standard and nonstandard loop quantum cosmology is described.
\end{abstract}

\pacs{...} \maketitle

\section{Introduction} \label{sec:intro}

The standard loop quantum cosmology (LQC) follows the Dirac
quantization scheme. In this approach, one first defines the {\it
kinematical} Hilbert space.  Then, the physical states are
determined by the requirement that the quantum constraint operator
(in simple cases there may be only one  constraint) vanishes on
them. The space of solutions is used to construct the {\it
physical} Hilbert space \cite{Ashtekar:2003hd,Bojowald:2006da}.
There exists an alternative method, the reduced phase space
approach, that we call nonstandard LQC. It consists in solving the
dynamical constraint already at the classical level and the
identification of physical observables. Examination of spectra of
the corresponding quantum observables leads to description of the
cosmological system. Such an approach has been recently applied to
the quantization of a flat Friedmann-Robertson-Walker (FRW) model
with a free massles scalar field
\cite{Dzierzak:2009ip,Malkiewicz:2009zd,Malkiewicz:2009qv}. In the
present  paper we consider an extension of the {\it classical}
aspects of the model by including a contribution from the
cosmological constant $\Lambda$.

Since the standard LQC seems to be very successful
\cite{Ashtekar:2010qn}, one may wonder what the {\it motivation}
for developing the nonstandard LQC could be. Let us discuss it in
more detail. The situation is that there are no {\it precise}
observational data available to verify the predictions of quantum
cosmology models. In such case a reasonable strategy seems to be
comparing results obtained within {\it alternative} approaches. An
agreement of results would prove that the {\it procedure} of
quantization is correct. An agreement of such results with
observational data (when they become available) would be the final
goal. The above strategy underlies our paper. We wish to obtain
results to be compared with the standard LQC results. On the other
hand, an alternative method may improve our understanding of
various conceptual issues like identification of Dirac's
observables, determination of the {\it minimum} length specifying
critical energy density of matter field at the big bounce
transition, quantum {\it evolution} of a system with Hamiltonian
constraint, etc. Present paper addresses the first issue and
begins the discussion of the second and third ones. They will be
considered in our next paper presenting quantization of the
present model.

\section{Hamiltonian}

The gravitational part of the classical Hamiltonian, in the
Ashtekar variables $(A^i_a,E^a_i)$, is the sum of the first class
constraints
\begin{equation}
H_{g} = \frac{1}{16\pi G} \int_{\Sigma} d^3{\bf x} (N^iC_i+N^aC_a+NC),
\end{equation}
where $\Sigma$ is the space-like part of spacetime $\dR \times
\Sigma$, and where
\begin{eqnarray}
C_i &=& D_a  E^a_i = \partial_a E^a_i+\epsilon_{ijk} A^j_a E^a_k, \\
C_a &=& E^b_iF^i_{ab}-(1+\gamma^2)K^i_aC_i, \\
C &=& \frac{ E^a_i E^b_j }{\sqrt{|\det E|}} \left( \epsilon^{ij}_k
F^k_{ab} -2(1+\gamma^2)  K^i_{[a}K^j_{b]} \right)
\end{eqnarray}
($\gamma$ is the Barbero-Immirzi parameter). For the considered
FRW model, the Gauss constraint $C_i$ as well as the spatial
diffeomorphisms constraint $C_a$ are automatically fulfilled. The
only nontrivial part is the scalar constraint $C$. Because for the
homogeneous models $K^i_a =\frac{1}{\gamma} A^i_a$, the
Hamiltonian simplifies to
\begin{equation}
H_{g} = -\frac{1}{\gamma^2}\frac{1}{16\pi G} \int_{\Sigma}
d^3{\bf x}\frac{1}{\sqrt{|\det E|}} E^a_i E^b_j \epsilon^{ij}_k F^k_{ab}. \label{Ham0}
\end{equation}

In LQC the gravitational degrees of freedom are parametrised by
holonomies $h_i$ and fluxes $F_i$ (which are functionals of the
Ashtekar variables). These are non-local functions used to
construct a non-perturbative  theory. The holonomies and fluxes
are non-trivial $SU(2)$ variables satisfying the holonomy-flux
algebra. However, in the highly symmetric spaces, like the FRW
model considered here, the forms of these functions simplify.

In particular, in the flat FRW model the flux may be
parametrised by   $v$ variable and the holonomy is expressed in
terms of $\beta$ variable \cite{Malkiewicz:2009qv}. The variable  $v$  is a physical
{\it volume} defined as follows
\begin{eqnarray}
    v  &:=& \int_\mathcal{V} dx_1 dx_2 dx_3 \sqrt {det \,q_{ab}}\nonumber \\
    &=&
    a^3 \int_\mathcal{V} dx_1 dx_2 dx_3 \sqrt {det \,q_{ab}^0}
    =: a^3\, V_0, \label{v1}
\end{eqnarray}
where $\mathcal{V}\subset \Sigma$ is an elementary {\it cell} in
the space with topology $\dR^3$; $(x_a) = (x^a) = (x^1, x^2, x^3)$
are Cartesian coordinates; $q_{ab}:= a^2 \,q_{ab}^0$ is a physical
3-metric; $a$ is a scale factor; $q_{ab}^0 dx^a dx^b:= dx_1^2 +
dx_2^2 + dx_3^2$ defines a fiducial 3-metric; $V_0$ is a fiducial
volume (it does not occur in final results). The $\beta$ variable,
in the limit $\beta \rightarrow 0$, is linked to the Hubble factor
$H =\dot{a}/a$ via the relation $\beta = \gamma H$.

In order to express the Hamiltonian, Eq. (\ref{Ham0}), in terms of
holonomies and fluxes, the procedure of regularization has to be
applied.  The regularization introduces a new scale to the theory,
namely the parameter $\lambda$. This can be understood as the
length scale of the lattice discretization. The applied procedure
of regularization and rewriting the Hamiltonian in terms of
holonomies and fluxes is the same as known from the standard LQC.
However, to the completeness of considerations we sum up the main
step of this derivation in Appendix A. The obtained gravitational
part of the Hamiltonian reads \cite{Ashtekar:2006wn}
\begin{eqnarray}
H^{(\lambda)}_g =  -\frac{vN}{32\pi^2 G^2\gamma^3\lambda^3} \sum_{ijk}
\epsilon^{ijk} \text{tr} \left[h_{\Box_{ij}}h_k \left\{ (h_k)^{-1}
,v\right\}\right], \nonumber  \\ \label{hgl}
\end{eqnarray}
where $h_{\Box_{ij}} = h_i h_j (h_i)^{-1} (h_j)^{-1}$ is the
holonomy around the square loop $\Box_{ij}$ (for more details see
\cite{Dzierzak:2009ip}). $N$ is the lapse function.  The
elementary holonomy in the i-th direction reads
\begin{equation}
h_i = \cos\left(\frac{\lambda
\beta}{2}\right)\mathbb{I}+2\sin\left( \frac{\lambda
\beta}{2}\right) \tau_i \label{hol}
\end{equation}
where $\tau_i=-\frac{i}{2} \sigma_i$ ($\sigma_i$ are the Pauli
matrices). The holonomy (\ref{hol}) is calculated in the
fundamental, $j=1/2$, representation of $SU(2)$. The factor
$\lambda$ is the parameter of the theory that may be related with
the minimum area of the loop. Namely,
$\text{Ar}_{\Box_{ij}}=\lambda^2$ so $\lambda$ has the dimension
of length. It is supposed to mark the scale when classical
dynamics should be modified by quantum effects. It is expected
that $\lambda \sim l_{\text{Pl}}$, but its precise value has to be
fixed observationally. At present, one can only give the upper
constraint on $\lambda$. Based on the observations of the cosmic
microwave background radiation it was recently shown
\cite{Mielczarek:2009zw} that $\lambda \leq 7 \cdot 10^4\,
l_{\text{Pl}}$.

In the model considered in this paper, the total Hamiltonian is
the sum of the gravity $H^{(\lambda)}_g$, cosmological constant
$H_{\Lambda}$, and  free scalar field  $H_{\phi}$ parts
\begin{equation}\label{hamtot}
H^{(\lambda)} =: N C := H^{(\lambda)}_g + H_{\Lambda} + H_{\phi},
\end{equation}
where
\begin{equation}\label{ham2}
H_{\Lambda} := N\frac{v\Lambda}{8\pi G}~~~~ \text{and}~~~~
H_{\phi} := N\frac{p^2_{\phi}}{2 v}.
\end{equation}

The insertion of the elementary holonomy (\ref{hol}) into Eq.
(\ref{hgl}) leads (for details, see Appendix A) to the expression
\begin{equation}
H^{(\lambda)} = N \left[-\frac{3}{8\pi G \gamma^2}
\frac{\sin^2(\lambda \beta) }{\lambda^2} v +\frac{p^2_{\phi}}{2 v}
+v \frac{\Lambda}{8\pi G} \right]  \approx 0.\label{Hamiltonian1}
\end{equation}
The variables $(v,\beta,\phi,p_{\phi})$ parametrise the phase
space, and the sign ``$\,\approx\,$'' reminds that the Hamiltonian
is a {\it constraint} of the gravitational system under
consideration.

The technical procedure  leading  to Eqs. (\ref{hgl}) and
(\ref{Hamiltonian1}) (presenting a Hamiltonian parametrized by
$\lambda$) is {\it identical} in both standard and nonstandard
LQC. The Hamiltonian must be the same, otherwise the comparison of
both methods could not be done. The real difference arises when
one begins the {\it implementation} of the Hamiltonian constraint,
$H^{(\lambda)} \approx 0$. In the standard approach, one promotes
Eq. (\ref{Hamiltonian1}) to an operator equation at the quantum
level with  the parameter $\lambda$ different from zero. Why is
the parameter $\lambda$ kept {\it different} from zero? The
technical answer is that in the limit $\lambda \rightarrow 0 $ the
{\it regular} constraint equation turns into the {\it singular}
Wheeler-DeWitt equation \cite{Dzierzak:2008dy}. The physical
justification for keeping $\lambda \neq 0$ offered by  the
standard method is that in the {\it loop} representation used to
quantize the kinematical level  this representation does not exist
for $\lambda = 0$ \cite{Ashtekar:2003hd}. In the second step of
this method, one turns the classical constraint equation into an
operator equation, which one defines consequently in the {\it
loop} representation of the kinematical level
\cite{Ashtekar:2006wn}. This is why one keeps $\lambda \neq 0$ in
the operator equation.  Why one uses such an exotic representation
which is not defined for $\lambda \neq 0 ?$ Further explanation is
that such representation is an analog of the representation used
at the kinematical level of loop quantum {\it gravity} (LQG)
\cite{Ashtekar:2003hd}. Is this answer satisfactory? It is
commonly known that the LQG has not been constructed yet. The
representation of the constraints algebra, based on the
achievements of the kinematical level, has not been found yet. The
problem is extremely difficult because the algebra is not a Lie,
but a Poisson algebra (one structure function is not a constant
but a function on phase space) \cite{TT}. The users of the
standard method believe that sooner or later this problem of LQG
will be solved and LQC will be derived from LQG, so making use of
an analogy to LQG is a healthy approach \cite{Ashtekar:2004eh}. We
think, it is an outstanding development, but far from being
completed.

We are conscious that in the standard LQC the parameter $\lambda$
is fixed by taking into account the spectrum of the kinematical
area operator of LQG (see, e.g. \cite{Ashtekar:2006wn}). Our
approach, making strong reference to observational cosmology, may
be treated as a sort of generalization of the standard approach.
It is not an effective semi-classical version of the standard LQC,
but a {\it modified} version of the classical cosmology  model of
general relativity. One can treat it as a one-parameter family of
classical Hamiltonians,  including the usual general relativity
Hamiltonian as a special case for $\lambda = 0$. It can be seen
easily that the singularity becomes `resolved' at the classical
level, for $\lambda \neq 0$, due to the functional form of Eq.
(\ref{Hamiltonian1}). It has been already discussed that the
regularization, making use of approximating the curvature of
connection by a holonomy around a loop with a finite length
$\sim\lambda$ (see, appendix A) produces the big bounce
\cite{Dzierzak:2009ip,Malkiewicz:2009qv,Haro:2009pt} already at
the classical level. Why should we quantize a cosmological model
which is free from the cosmological singularity? There are at
least three good reasons: (i) we must have a quantum model to make
comparison with the standard method results which concern {\it
quantum} level, (ii) classical energy density of matter depends on
a free parameter $\lambda$ in such a way that it may become
arbitrary big for small enough $\lambda$ (see Eqs. (\ref{den1}),
(\ref{rhods}) and (\ref{den2})) so the system may enter a length
scale where quantum effects have to be taken into account, and
(iii) making predictions of our model for quantum cosmic data may
be used to {\it fix} the free parameter $\lambda$, after such data
become available.

The model with the Hamiltonian defined by Eq. (\ref{Hamiltonian1})
has been already studied \cite{Mielczarek:2008zv} and analytical
solutions have been found:

\noindent The Hamiltonian constraint,  Eq. (\ref{Hamiltonian1}),
can be rewritten as
\begin{equation}
\frac{2v^2}{p^2_{\phi}} = \frac{1}{\rho_{\text{c}}}
\frac{1}{\left[\sin^2 (\lambda \beta)-\delta\right]} \geq 0
\label{vp},
\end{equation}
where $\rho_{\text{c}}$ is the critical energy density
\begin{equation}\label{den2}
\rho_{\text{c}} := \frac{3}{8\pi G \gamma^2 \lambda^2}
\end{equation}
and $\delta$ is a parameter
\begin{equation}
\delta :=\frac{\rho_{\Lambda}}{\rho_{\text{c}}} =
\frac{\Lambda}{3} \gamma^2 \lambda^2 , \label{deltadef}
\end{equation}
and where  $\rho_{\Lambda} := \frac{\Lambda}{8\pi G}$.

It turns out  that for $ \rho_{\Lambda} > \rho_{\text{c}}\ (\delta
> 1)$, the system has no physical solutions \cite{Mielczarek:2008zv}.
The physical
solutions exist only for $\Lambda<\Lambda_{\text{c}}$, where
\begin{equation}
\Lambda_{\text{c}} := 8\pi G \rho_{\text{c}} =
\frac{3}{\gamma^2\lambda^2},
\end{equation}
which defines the critical value of the cosmological constant.
This surprising result extends to the quantum level as well
\cite{Bentivegna:2008bg,Kaminski:2009pc}.

In what follows we consider two cases: $\Lambda<0$ ($\delta<0$,
anti-de Sitter) and $0<\Lambda<\Lambda_{\text{c}}$ ($0<\delta<1$,
de Sitter).

\section{Equations of motion}

The equations of motion for the system are defined by the Hamilton
equation $\dot{f}=\left\{f,H^{(\lambda)} \right\}$, where the
Poisson bracket is defined as follows
\begin{eqnarray}
\{ \cdot , \cdot \} &:=& 4\pi G \gamma \left[\frac{\partial \cdot }
{\partial \beta} \frac{\partial \cdot }{\partial v}-
\frac{\partial \cdot }{\partial v}\frac{\partial \cdot }{\partial \beta}
\right]\nonumber \\
 &+& \frac{\partial \cdot }{\partial \phi} \frac{\partial \cdot }
 {\partial p_{\phi}}-
\frac{\partial \cdot }{\partial p_{\phi}}\frac{\partial \cdot }
{\partial \phi}. \label{Poisson1}
\end{eqnarray}
The solutions of the Hamilton equation with $f \in
(v,\beta,\phi,p_{\phi})$ define the kinematical phase space
$\mathcal{F}_{\text{kin}}$. In turn, the solutions restricted by
the Hamiltonian constraint, Eq. (\ref{Hamiltonian1}), define the
physical phase space $\mathcal{F}_{\text{phys}}$.

For the function defined on $\mathcal{F}_{\text{phys}}$, the
dynamics are governed by
\begin{equation}
\dot{f} = \left\{f,NC\right\} = \underbrace{\left\{f,N\right\}C}_{
= \;0 \ \text{on} \ \mathcal{F}_{\text{phys}}
}+N\left\{f,C\right\} = N\left\{f,C\right\}. \label{fdot}
\end{equation}
Thus, the {\it relational} dynamics of variables $f$ and $g$ on
$\mathcal{F}_{\text{phys}}$ is defined by
\begin{equation}
\frac{\dot{f}}{\dot{g}} = \frac{df}{dg} = \frac{
\left\{f,C\right\} }{ \left\{g,C\right\} }
\end{equation}
so it is gauge independent. The specific choice of gauge can
however simplify the calculations. In our considerations we choose
the lapse function in the form
\begin{equation}
 N  = 1/v .
\end{equation}

The equations of motion read
\begin{eqnarray}
\dot{v} &=&  \{v,H^{(\lambda)}\} = \frac{3 \sin(2 \beta  \lambda )}
{2 \gamma  \lambda } \label{eom1}, \\
\dot{\beta} &=&  \{\beta,H^{(\lambda)}\} = -\frac{4 G p^2_{\phi} \pi
\gamma }{v^3} \label{eom2},\\
\dot{\phi} &=&  \{\phi,H^{(\lambda)}\} = \frac{p_{\phi}}{v^2}  \label{eom3}, \\
\dot{p}_{\phi} &=& \{p_{\phi},H^{(\lambda)}\} = 0 \label{eom4}.
\end{eqnarray}

An {\it elementary} observable $\mathcal{O}$ is a real function on
the phase space that satisfies the equation (for a precise
definition see \cite{Dzierzak:2009ip})
\begin{equation}\label{con}
\{\mathcal{O},H^{(\lambda)}\} \approx 0.
\end{equation}
It is clear that constants of motion of the Hamilton equations
satisfy Eq. (\ref{con}).  We get immediately from Eq. (\ref{eom4}) the
first observable
\begin{equation}
\mathcal{O}_1 = p_{\phi}.
\end{equation}
Finding other observables needs definitely more effort.

Before  we proceed to the task we would like to firstly make a
comment on the relation between the classical cosmology and the
model under considerations. In particular, we would like to show
in which limit, the standard Friedmann equation is recovered.

We define the Hubble factor as follows $H:= \frac{1}{3v}
\frac{dv}{dt}$, where $t$ is a coordinate time.  We can insert Eq.
(\ref{eom1}) into this definition, however keeping in mind the
fact that $\dot{v}=\frac{1}{N}\frac{dv}{dt}$, which results from
Eq. (\ref{fdot}). Taking square of the Hubble factor and inserting
Eq. (\ref{vp}) we find
\begin{equation}
H^2=\frac{8 \pi G}{3} \rho_{\phi} \left(1-\frac{\rho_{\phi}}{\rho_{\text{c}}}\right)
+\frac{\Lambda}{3}\left(1-2\frac{\rho_{\phi}}{\rho_{\text{c}}}-\frac{\rho_{\Lambda}}
{\rho_{\text{c}}} \right), \label{ModFried}
\end{equation}
where we have defined $\rho_{\phi} := \frac{p^2_{\phi}}{2v^2}$.
Equation (\ref{ModFried}) is the modified Friedmann equation which
results from the considered model. Let us consider the limit
$\lambda \rightarrow 0$. Based on Eq. (\ref{den2}) we find
$\lim_{\lambda \rightarrow 0}\rho_{\text{c}}=\infty$. Therefore in
the limit of vanishing modification,  $\lambda \rightarrow 0$, the
classical Friedmann equation
\begin{equation}
H^2=\frac{8 \pi G}{3} \rho_{\phi} +\frac{\Lambda}{3}, \label{ClasFried}
\end{equation}
is recovered. Another issue is the \emph{correspondence} with the
standard cosmology. We see that the modifications become
irrelevant when $\rho_{\phi} \ll \rho_{\text{c}}$ and
$\rho_{\Lambda} \ll \rho_{\text{c}}$, and the form of Eq.
(\ref{ClasFried}) is recovered. Therefore, the correspondence
takes place in the limit of low energy densities of matter and
cosmological constant. However, while the matter energy density
$\rho_{\phi}$ is diluted with the increase of volume $v$, the
$\rho_{\Lambda}$ remains constant. Therefore the last term in Eq.
(\ref{ModFried}), namely $-\frac{\Lambda}{3}
\frac{\rho_{\Lambda}}{\rho_{\text{c}}} =-\frac{\Lambda}{3}\delta
$, contributes also in the limit of the large volumes. As we will
show later, the value of parameter $\delta$ is extremely low for
the observationally determined value of cosmological constant.
This additional constant term is therefore completely negligible
and its present contribution cannot be verified observationally.

\section{Anti-de Sitter (AdS)}

Based on Eqs. (\ref{eom1}) and (\ref{eom2}) we find
\begin{equation}
\frac{d\beta}{dv} = -\frac{8\pi G p^2_{\phi}\gamma^2 \lambda}
{3 \sin(2 \beta  \lambda )v^3}.
\end{equation}
Integrating the above equation we obtain the observable (as a
constant of integration)
\begin{equation}
\mathcal{O}_3 = \cos(2\beta \lambda)+\frac{8\pi G \gamma^2 \lambda^2}
{3v^2}\mathcal{O}_1^2.
\end{equation}
Equations (\ref{eom2}) and (\ref{eom3}) give
\begin{equation}
\frac{d\beta}{d\phi} = -\frac{4\pi G p_{\phi}\gamma}{v}. \label{dbdphi}
\end{equation}
Solving this equation gives the next observable
\begin{eqnarray}
\mathcal{O}_2 &=&\phi+ \frac{2}{\sqrt{8\pi G}} \frac{\text{sgn}(p_{\phi})}
{\sqrt{3(\mathcal{O}_3-1)}}
\left[ F\left(\beta\lambda \left|-\frac{2}{\mathcal{O}_3-1} \right.\right)
\right. \nonumber \\
&-&\left.  F\left(\frac{\pi}{2} \left|-\frac{2}{\mathcal{O}_3-1} \right.\right)
\right] \label{O21}
\end{eqnarray}
where $F$ is the Jacobi elliptic integral of the first kind
\begin{equation}
F(x|m) =\int_0^x\frac{d\theta}{\sqrt{1-m\sin^2(\theta)}}. \label{Jacobi}
\end{equation}
To get Eq. (\ref{O21}) we have spitted the integration of the
righthand side of Eq. (\ref{Jacobi}) as follows
\begin{eqnarray}
\int_{\pi/2}^x \frac{d\theta}{\sqrt{1-m\sin^2\theta}} &=&
\int_{0}^x \frac{d\theta}{\sqrt{1-m\sin^2\theta}}  \nonumber \\
&-&\int_0^{\pi/2}\frac{d\theta}{\sqrt{1-m\sin^2\theta}} \label{integrals}.
\end{eqnarray}
The lower limit of the integration, in variable
$\theta:=\lambda\beta$, is equal to $\pi/2$. With this particular
choice, the limit $\Lambda \rightarrow 0$ leads to the expressions
found in the case $\Lambda=0$. This limit will be discussed in
more details in Sec. \ref{Limit}. In fact, the lower limit of
integration can be set to be an arbitrary number. In particular,
it can equals not $\pi/2$, but zero. This will not change the
physics, but it will not lead to  the  expression for the
observables found in the case $\Lambda \rightarrow 0$. It is so
because the elementary observables are defined up to an additive
constant.

One may verify that obtained observables satisfy the Lie algebra
\begin{eqnarray}
\left\{ \mathcal{O}_1,\mathcal{O}_3 \right\} &=& 0,  \\
\left\{ \mathcal{O}_2,\mathcal{O}_3 \right\} &=& 0,  \\
\left\{ \mathcal{O}_2,\mathcal{O}_1 \right\} &=& 1.
\end{eqnarray}
This algebra can be simplified since the observable
$\mathcal{O}_{3}$ can be eliminated due to the Hamiltonian
constraint which reads
\begin{equation}\label{con2}
\mathcal{O}_{3}=1-2\delta.
\end{equation}
Making use of Eq. (\ref{con2}) we may  rewrite  $\mathcal{O}_{2}$ in
the form
\begin{equation}
\mathcal{O}_2 = \phi+ \frac{\text{sgn}(p_{\phi})}{\sqrt{12\pi
G(-\delta)}} \left[F\left(\beta\lambda \left|\frac{1}{\delta} \right.
\right)-K\left( \frac{1}{\delta}\right)\right]\label{O2}
\end{equation}
where $K\left(m\right)=F\left(\pi/2\left| m \right.\right)$.
It Appendix B, the form of the Poisson bracket on
$\mathcal{F}_{\text{phys}}$ has been derived. The observables
$\mathcal{O}_1$ and $\mathcal{O}_2$ satisfy on
$\mathcal{F}_{\text{phys}}$  the Lie algebra
\begin{equation}
\left\{ \mathcal{O}_2,\mathcal{O}_1 \right\} = 1 ,
\end{equation}
where
\begin{equation}
 \left\{ \cdot , \cdot \right\} :=
\frac{\partial \cdot}{\partial \mathcal{O}_2} \frac{\partial \cdot }
{\partial \mathcal{O}_1} -
\frac{\partial \cdot }{\partial \mathcal{O}_1} \frac{\partial \cdot }
{\partial \mathcal{O}_2}.
\end{equation}

Equation (\ref{O2}) can be inverted to the form
\begin{equation}
\lambda\beta = \text{am}\left(\sqrt{12\pi G
(-\delta)}\text{sgn}(p_{\phi}) (\mathcal{O}_2-\phi)+K\left(\frac{1}
{\delta}\right)\left|
\frac{1}{\delta} \right. \right), \label{lb}
\end{equation}
where am$(x|m) = F^{-1}(x|m)$ in an amplitude of the Jacobi integral.

\begin{figure}[ht!]
\centering
\includegraphics[width=7cm,angle=0]{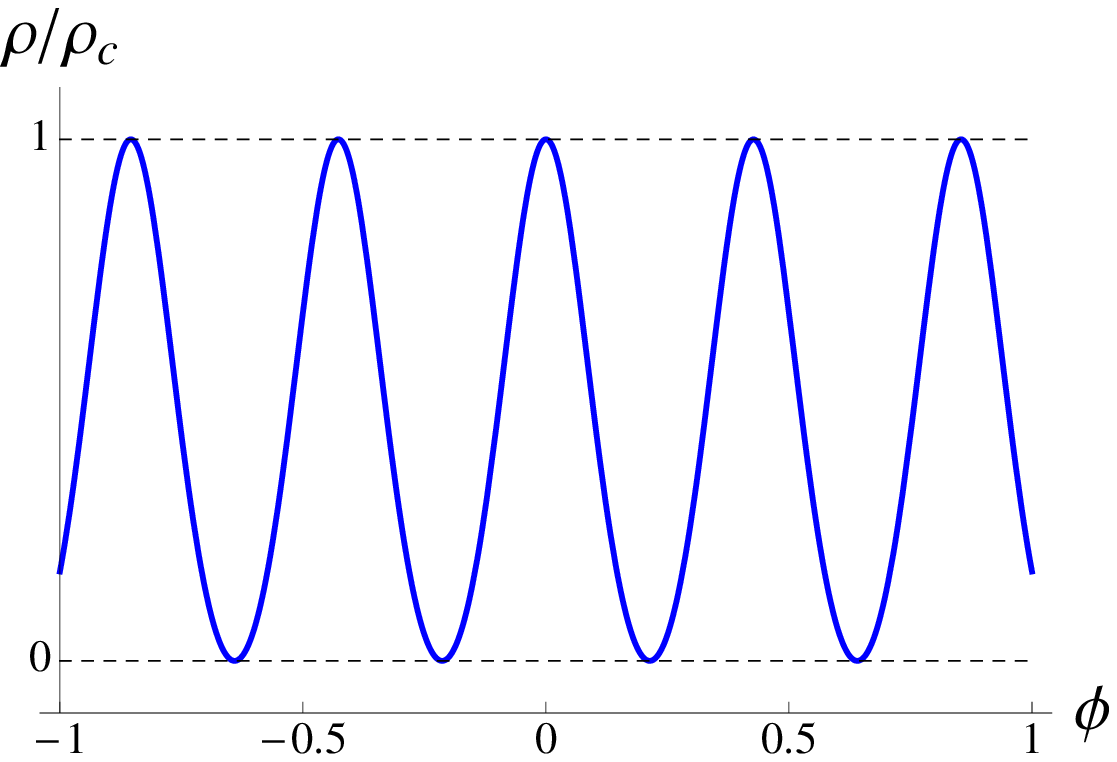}
\caption{Plot of $\rho(\phi)$ for $\delta=-1$ and
$\mathcal{O}_2=0$. Here, the period of oscillations is equal to
$\Delta\phi\simeq 0.427$} \label{rAdS}
\end{figure}

The energy density is found to be
\begin{equation}
\rho=\rho_{\phi}+\rho_{\Lambda} = \frac{p^2_{\phi}}{2v^2} +\frac{\Lambda}
{8\pi G}. \label{den1}
\end{equation}
Due to  the Hamiltonian constraint it reads
\begin{equation}\label{den}
\rho=\rho_{\text{c}}\, \sin^2(\lambda\beta).
\end{equation}
Using Eq. (\ref{lb}) we may rewrite Eq. (\ref{den}) in the form
\begin{equation}\label{den1}
\rho(\phi)=\rho_{\text{c}} \text{sn}^2\left( \sqrt{12\pi
G(-\delta)}
(\mathcal{O}_2-\phi)+K\left(\frac{1}{\delta}\right)\left|
\frac{1}{\delta} \right. \right),
\end{equation}
where $\text{sn}(\cdot|\cdot):=\sin(\text{am}(\cdot|\cdot))$. We
plot this dependence in Fig. \ref{rAdS}.

\begin{figure}[ht!]
\centering
\includegraphics[width=7cm,angle=0]{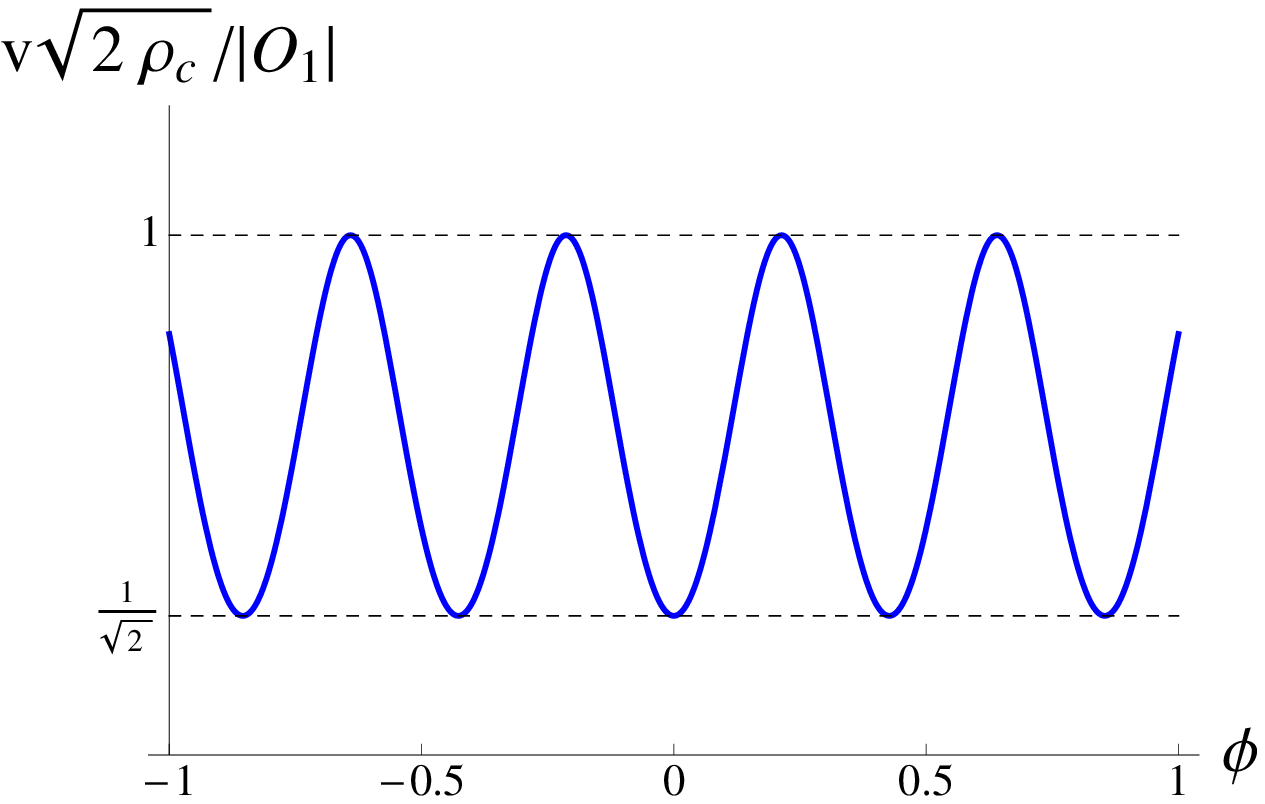}
\caption{Plot of $v(\phi)$ for $\delta=-1$ and $\mathcal{O}_2=0$.
Here, the period
of oscillations is equal to $\Delta\phi\simeq 0.427$}
\label{vAdS}
\end{figure}

Based on Eqs. (\ref{vp}) and  (\ref{lb}) we derive an expression
for the volume observable in terms of elementary observables and
an evolution parameter $\phi$
\begin{equation}
v (\phi) = \frac{|\mathcal{O}_1|}{\sqrt{2\rho_c}}
\frac{1}{\sqrt{\text{sn}^2\left( \sqrt{12\pi G (-\delta)}
(\mathcal{O}_2-\phi)+K\left(\frac{1}{\delta}\right)\left| \frac{1}{\delta} \right.
\right)-\delta}}.
\end{equation}
The plot of this function is shown in Fig. \ref{vAdS}.
The solution for the volume $v$ are non-singular oscillations with
the maximal and minimal values given by
\begin{eqnarray}
v_{\text{max}}  &=&  \frac{|\mathcal{O}_1|}{\sqrt{2\rho_c}}  \frac{1}
{\sqrt{-\delta}}   \ \  \mapsto \ \ \rho = 0, \label{vmax}    \\
v_{\text{min}}  &=&  \frac{|\mathcal{O}_1|}{\sqrt{2\rho_c}}  \frac{1}
{\sqrt{1-\delta}}  \ \  \mapsto \ \ \rho = \rho_{\text{c}}. \label{vmin}
\end{eqnarray}

Equation (\ref{vmax}) tells that when the volume is maximal, the
total energy density equals zero. However, the matter density does
not vanish, since due to Eq. (\ref{den1})  we have
$\rho_{\phi}=-\rho_{\Lambda}=-\frac{\Lambda}{8\pi G} > 0$ (as
$\Lambda <0$ in  AdS case).

The period of oscillations can be written as
\begin{equation}
\Delta \phi = \frac{2K\left( \frac{1}{\delta} \right)}{\sqrt{12\pi G (-\delta)}}.
\end{equation}

\section{de Sitter (dS)}

In this case the considerations are similar to those presented in
the previous section. However, finding a suitable definition of
the observable $\mathcal{O}_2$ requires some additional analysis.
Namely, in this case
\begin{equation}
\sin^2(\lambda\beta) \geq \delta
\end{equation}
that result from Eq. (\ref{vp}). Thus, $\lambda\beta \in
[\theta_0+n\pi,(1-n)\pi-\theta_0]$, where $n \in \mathbb{Z}$ and
$\theta_0 := \arcsin(\sqrt{\delta})$. The equation for the
observable $\mathcal{O}_2$ takes the form
\begin{equation}
d\mathcal{O}_2 =d\phi+\frac{\text{sgn}(p_{\phi})}{\sqrt{12\pi G}}
\frac{\lambda d\beta}{\sqrt{\sin^2(\lambda \beta)-\delta}}. \label{dO2}
\end{equation}
The integration of this equation can be done by using the Jacobi
function (see Eq. (\ref{Jacobi})) as in anti-de Sitter case. The
lower limit of integration is set to be $\pi/2$ and the relation
(\ref{integrals}) is applied. The choice of the lower limit is
dictated by the limit $\Lambda \rightarrow 0$. For the
$\theta<\theta_0$ the integration gives an imaginary number.
However, due to relation (\ref{integrals}) these contributions
cancel out. Finally, we get
\begin{equation}
\mathcal{O}_2 =\phi+\frac{\text{sgn}(p_{\phi})}{\sqrt{12\pi G
\delta}} \frac{1}{i} \left[ F\left( \beta\lambda \left|
\frac{1}{\delta} \right. \right)-K\left(\frac{1}{\delta}\right)\right], \label{O2dS}
\end{equation}
which is a real function on $\mathcal{F}_{\text{phys}}$ for $\lambda\beta \in
[\theta_0,\pi-\theta_0]$. The algebra of observables is the same as in the
anti-de Sitter case.

\begin{figure}[ht!]
\centering
\includegraphics[width=7cm,angle=0]{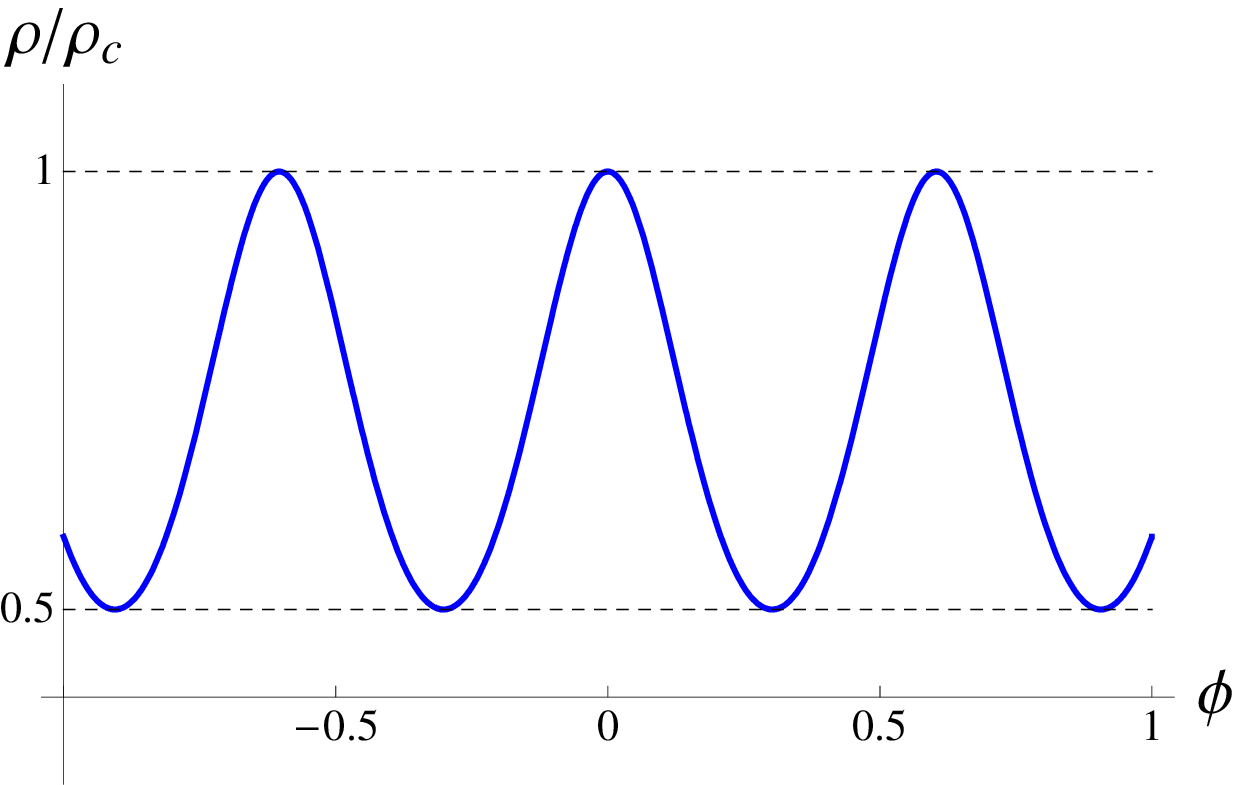}
\caption{Plot of $\rho(\phi)$ for $\delta=0.5$ and $\mathcal{O}_2=0$.
Here, $\Delta\phi \simeq 0.604$.}
\label{rdS}
\end{figure}

Equation (\ref{O2dS}) can be inverted to the form
\begin{equation}
\sin^2(\lambda\beta)=\delta \text{dn}^2 \left(\sqrt{12\pi G \delta}
(\mathcal{O}_2-\phi)+K\left|1-\frac{1}{\delta}  \right. \right)
\label{sin2}
\end{equation}
where we defined $K:=K\left(1-\frac{1}{\delta}\right)$, so we have
\begin{equation}
\rho(\phi)= \rho_{\text{c}}\delta \text{dn}^2 \left(\sqrt{12\pi G \delta}
(\mathcal{O}_2-\phi)+K \left|1-\frac{1}{\delta}  \right. \right). \label{rhods}
\end{equation}
The function $\text{dn}(\cdot|\cdot)$ is defined as
$\text{dn}(u|m) :=\sqrt{1-m\sin^2[\text{am}(u|m)]}$. We plot
function (\ref{rhods}) in Fig. \ref{rdS}.

\begin{figure}[ht!]
\centering
\includegraphics[width=7cm,angle=0]{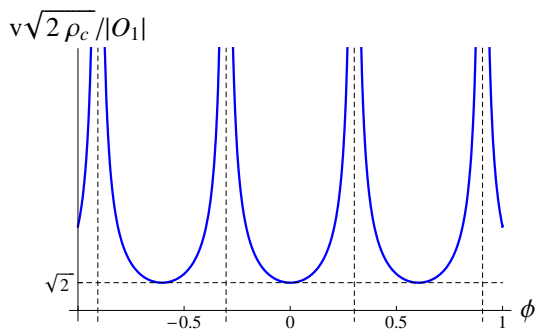}
\caption{Plot of $v(\phi)$ for $\delta=0.5$ and $\mathcal{O}_2=0$.
Here, $\Delta\phi \simeq 0.604$.}
\label{vdS}
\end{figure}

Making use of Eqs. (\ref{sin2}) and  (\ref{vp}) we find
\begin{equation}
v(\phi) = \frac{|\mathcal{O}_1|}{\sqrt{2\rho_c(1-\delta)}}
\frac{1}{\left|\text{sn}\left(\sqrt{12\pi G \delta}(\mathcal{O}_2-\phi)+K
\left|1-\frac{1}{\delta}  \right. \right)  \right|}. \label{vds}
\end{equation}
We present this function in Fig. \ref{vdS}.
The solution is periodic
\begin{equation}
v(\phi)=v(\phi+n\,\Delta\phi) ,
\end{equation}
where $n \in \mathbb{Z}$ and
\begin{equation}
\Delta\phi :=\frac{2K\left( 1-\frac{1}{\delta}
\right)}{\sqrt{12\pi G \delta}}.
\end{equation}
However,  contrary to the Anti-de Sitter case, the periodicity
does not mean that the volume oscillates. The cycles are separated
by asymptotes with $v\rightarrow \infty$. Each cycle is called a
bounce. The bounces are disconnected and define quite independent
solutions. Within a bounce, the value of the scalar field change
by $\Delta\phi$. One can find that this corresponds to the
evolution of the system from $t=-\infty\,$ to $\,t=+\infty$, where
$t$ in the coordinate time. Thus, each period corresponds to the
same  dS universe.

It turns out \cite{ Kaminski:2009pc,WK} that in the standard LQC
the quantum evolution of semi-classical states interpolate between
the trajectories in the adjacent eras shown  in Fig. 4, forming a
cyclic system. We plan to examine this interesting issue, within
the nonstandard LQC, in our next paper concerning a quantum
version of the model.

The maximum and minimum of of the volume, and the corresponding
energy densities are
\begin{eqnarray}
v_{\text{max}}  &=&   \infty  \ \  \mapsto \ \ \rho = \rho_{\Lambda},   \\
v_{\text{min}}  &=&   \frac{|\mathcal{O}_1|}{\sqrt{2\rho_c}}\frac{1}{\sqrt{1-\delta}}
\ \  \mapsto \ \ \rho = \rho_{\text{c}}.
\end{eqnarray}

It may appear to be strange that as the maximum volume equals
infinity, the total energy density does not vanish. However, this
counter-intuitive result is correct since due to Eq. (\ref{den1})
the matter term goes to zero as the volume approaches infinity,
whereas the $\Lambda$-term stays finite and equals $\Lambda/8\pi
G>0$ (as $\Lambda > 0 $ in  dS case).

\section{The limit $\delta \rightarrow 0$} \label{Limit}

In the case $\delta=0$, the observables take the following form
\begin{eqnarray}
\mathcal{O}_1 &=& p_{\phi}, \label{O1pphi} \\
\mathcal{O}_2 &=& \phi+ \frac{\text{sgn}(p_{\phi})}{\sqrt{12\pi G}}
\ln\left|  \tan\left(\frac{\beta\lambda}{2}\right)\right|.\label{O2phi} \
\end{eqnarray}
To find the observable Eq. (\ref{O2phi}), we have integrated Eq.
(\ref{dbdphi}) taking into account the Hamiltonian constraint with
$\delta=0$.

These observables satisfy the same Lie algebra as in the case
$\delta\neq 0$. Namely,
$\left\{\mathcal{O}_2,\mathcal{O}_1\right\}=1$. Making use of the
identity
\begin{equation}
\ln\left| \tan\left(\frac{\beta\lambda}{2}\right)\right| = -\text{arth}
\left[ \cos(\beta\lambda ) \right],
\end{equation}
one can find that the observables, Eqs. (\ref{O1pphi}) and
(\ref{O2phi}), are the same as those derived in
\cite{Dzierzak:2009ip}.

The elementary observables play the role of `building blocs' used
to define {\it composite} observables. In particular, one can find
an expression for the volume
\begin{equation}
v = \frac{|\mathcal{O}_1|}{\sqrt{2\rho_{\text{c}}}} \cosh \left[\sqrt{12\pi G }
(\mathcal{O}_2-\phi) \right], \label{v0}
\end{equation}
as well as for the energy density
\begin{equation}
\rho = \frac{\rho_{\text{c}}}{\cosh^2 \left[\sqrt{12\pi G }(\mathcal{O}_2-\phi)
\right]}. \label{rho0}
\end{equation}
These compound observables are gauge {\it independent} and overlap
with those found in \cite{Dzierzak:2009ip}.

Now, let us discuss the limit $\delta\rightarrow 0$, for fixed
$\lambda$. The observable $\mathcal{O}_1$ remains unchanged as it
is independent on $\delta$. Let us examine the case of the
observable $\mathcal{O}_2$. Here the situation for both anti-de
Sitter and se Sitter case looks the same. Equations (\ref{O2}) and
 (\ref{O2dS}) can be written as
\begin{equation}
\mathcal{O}_2 = \phi+\frac{\text{sgn}(p_{\phi})}{\sqrt{12\pi
G}} \int_{\pi/2}^{\lambda\beta} \frac{d\theta}{\sqrt{\sin^2\theta-\delta}}.
\end{equation}
We have
\begin{eqnarray}
\lim_{\delta\rightarrow 0^{\pm}} \int_{\pi/2}^x \frac{d\theta}{\sqrt{\sin^2\theta-\delta}}
= \int_{\pi/2}^x \frac{d\theta}{\sin \theta} = \ln \left| \tan\left(\frac{x}{2}\right) \right|
\end{eqnarray}
in the definition of $\mathcal{O}_2$. Thus, in both limits $\delta
\rightarrow 0^{\pm}$ the expression Eq. (\ref{O2phi}) is
recovered.

Now, let us try to get Eqs. (\ref{v0}) and (\ref{rho0}) by taking
the limit $\delta \rightarrow 0^{\pm}$. Firstly, let us consider
de Sitter's case. We rewrite equations (\ref{rhods}) and
(\ref{vds}) in the form
\begin{eqnarray}
\rho &=&\rho_{\text{c}}\left[\delta-(\delta-1)\sin^2\varphi\right], \label{limr} \\
v &=& \frac{|\mathcal{O}_1|}{\sqrt{2\rho_{\text{c}}(1-\delta)}}\frac{1}{|\sin \varphi|},
\label{limv}
\end{eqnarray}
where
\begin{equation}
\varphi = \text{am} \left[\sqrt{12\pi G \delta}(\mathcal{O}_2-\phi)+K
\left|1-\frac{1}{\delta}  \right. \right].
\end{equation}
The last equality can be  written as
\begin{equation}
\sqrt{12\pi G}(\mathcal{O}_2-\phi) = \int_{\pi/2}^{\varphi}\frac{d\theta}
{\sqrt{\delta-(\delta-1)\sin^2\theta}}.
\end{equation}
Taking the limit $\delta \rightarrow 0^+$, we get
\begin{equation}
\sqrt{12\pi G}(\mathcal{O}_2-\phi) = \int_{\pi/2}^{\varphi}\frac{d\theta}{\sin\theta} =
\ln \left| \tan\left(\frac{\varphi}{2}\right) \right|
\end{equation}
which can be rewritten as
\begin{equation}
\sin^2{\varphi} = \frac{1}{\cosh^2 \left[\sqrt{12\pi G }(\mathcal{O}_2-\phi)
\right]}.
\end{equation}
It is clear now, that in the limit $\delta \rightarrow 0^+$, Eq.
(\ref{limr}) and Eq. (\ref{limv}) turn into Eq. (\ref{rho0}) and
Eq. (\ref{v0}), respectively.

In the anti-de Sitter case the procedure is similar.
This way both limits, $\delta \rightarrow 0^{\pm}$, lead to the
same expressions for the density and the volume.

Now, let us determine the parameter $\delta$ from observations.
The cosmological constant $\Lambda$ can be related with the
observed dark energy, which dominates the energy density of the
Universe. In this case one can rewrite the definition Eq.
(\ref{deltadef}) in the form
\begin{equation}
\delta = \frac{1}{c^2} \Omega_{\Lambda} H^2_0 \gamma^2 \lambda^2,
\end{equation}
where $\Omega_{\Lambda}$ is the fractional density of the cosmological constant,
 $H_0$ is the present value of the Hubble parameter, and $c$ is the speed of light.
The five years observations of the WMAP satellite yield
$\Omega_{\Lambda}=0.742 \pm 0.030$ and $H_0=71.9^{+2.6}_{-2.7} \
\text{km} \ \text{s}^{-1}\ \text{Mpc}^{-1}$ \cite{Dunkley:2008ie}.
Assuming that $\lambda=l_{\text{Pl}}$, where $\l_{\text{Pl}}$ is
the Planck length and $\gamma=\gamma_M=0.2375\,$
\cite{Meissner:2004ju}, we find
\begin{equation}
\delta = 6.6 \cdot 10^{-124}
\end{equation}
which is an extremely small value. Such a small vale of $\delta$
is connected with the known discrepancy between observed value of
the cosmological constant and the energy density of the quantum
vacuum. Since $\delta>0$, the present observations favour dS case
rather than AdS. It is, however, not excluded that the AdS phase
had some realization in the past.

\section{Conclusion}

The density and the volume are functions of the elementary
observables and an evolution parameter $\phi$. They become
observables for each fixed value of this parameter as in such a
case they satisfy Eq. (\ref{con}). Due to Eq. (\ref{eom3}) the parameter
changes monotonically so it suits the purpose.

For any  value of the cosmological constant $\Lambda$, considered
in our paper,  the elementary observables satisfy the same simple
Lie algebra $\left\{ \mathcal{O}_2,\mathcal{O}_1 \right\} = 1$, on
the constraint surface. The elementary observables serve as
building blocks for composite ones. They correspond to the
constants of motion specifying dynamics so they have physical
meaning, but they are not required to be measurable.

The composite observables are defined on the physical phase space
and have clear physical interpretation, so they are expected to be
{\it detectable} in observational cosmology. The volume is bounded
and oscillates in the AdS case. It is bounded from below and
diverges in the dS case, as expected. The energy density is
bounded in both cases. We have shown that in the limit $\Lambda
\rightarrow 0^{\pm}$ the observables obtained for the case
$\Lambda=0$ are recovered. Thus, our results are consistent.

The initial big-bang singularity  {\it turns} into the big-bounce
transition. The density is a function of a {\it free} parameter
$\lambda$ and blows up as $\lambda \rightarrow 0$, which
corresponds to the case when there is no modification of the
Hamiltonian by the holonomies. In both standard and nonstandard
LQC  $\lambda$ is a {\it free} parameter. It seems there is no
satisfactory way to determine its theoreticall value, but
forthcoming observtional data may bring some resolution to this
problem. We have already addressed this issue in the context of
the standard LQC \cite{Dzierzak:2008dy} and in the nonstandard LQC
\cite{Dzierzak:2009ip,Malkiewicz:2009qv}. Some preliminary
agreement with our results (without reference to ours) may be
found in Sec. VI of an updated version 4 of
\cite{Ashtekar:2007em}. One discusses there `parachuting by hand'
of the results from full LQG into LQC, since the derivation of LQC
from LQG has not been obtained yet.

One usually relates the violation of the Lorentz symmetry with
quantum gravity effects. Recent observations suggest \cite{nature}
that the scale of the Lorentz symmetry violation is greater than
$1.2 E_{\text{Pl}}=1.5 \cdot 10^{19}\text{GeV}$, which corresponds
to the length scale $0.8 l_{\text{Pl}}$.  If quantum effects lead
to the violation of the Lorentz symmetry, then the length scale of
this effect should be related with $\lambda$.  In such a case the
astrophysical observations of the $\gamma-$ray bursts, like those
presented in \cite{nature}, can be potentially used to constrain
the parameter $\lambda$. However, an explicit functional form of
this relation  is unknown. On the other hand, the observations
like \cite{nature} are still ambiguous due to low statistics.
Thus, one cannot impose on $\lambda$ any realistic astrophysical
constraints yet.

Our method  relies on a direct link with observational data due to
the unknown value of $\lambda$. This may make our model suitable
for describing observational data despite the fact that FRW may
have too much symmetry to be a realistic model of the Universe.
Taking theoretically determined $\lambda$ from {\it incomplete}
LQG, in the standard LQC, seems to give a model of the Universe
less realistic than ours. We believe that lacking of theoretically
determined numerical value of $\lambda$ is rather meritorious than
problematic.

In our next paper we shall present {\it quntization} of the model
considered here. This will enable us making {\it complete} comparison
of the standard and the nonstandard LQC.\\

\begin{acknowledgments}
We thank Wojtek Kami\'{n}ski  for helpful correspondence. JM has
been supported by Polish Ministry of Science and Higher Education
grant N N203 386437.
\end{acknowledgments}

\appendix

\section{Modified Hamiltonian} \label{Appendix1}

In this  appendix we give more details on the modified
gravitational Hamiltonian used it this paper. In particular, we
derive the form of the regularized Hamiltonian, Eq. (\ref{hgl}).
Later, we show how this Hamiltonian simplifies to Eq.
(\ref{Hamiltonian1}) after the expression for the holonomy $h_i$
in applied. We begin from Eq.  (\ref{Ham0}). Applying the
classical identity
\begin{equation}
\frac{1}{\sqrt{|\det E|}}E^a_i E^b_j \epsilon^{ij}_k F^k_{ab} = \frac{1}{4\pi G
\gamma}\epsilon^{abc} \{A^i_c, v \} F_{abi},
\end{equation}
and the trace of a product of the $SU(2)$ variables we find
\begin{equation}
H_g = \frac{1}{32\pi^2 G^2}\frac{1}{\gamma^3} \int_{\Sigma}  d^3{\bf x} N
\epsilon^{abc} \text{tr}\left[F_{ab} \{A_c, v \} \right].
\label{ApAHg}
\end{equation}
Regularization of this Hamiltonian can be performed with use of expressions
\begin{equation}
{^o}e^a_k \{A_a,v\} \approx -\frac{1}{\mu V^{1/3}_0} h_k \{h^{-1}_k ,v\} \label{reg1}
\end{equation}
and
\begin{eqnarray}
h_{\Box_{ij}}    &=& h_i h_j (h_i)^{-1} (h_j)^{-1}    \nonumber \\
                        &\approx& \mathbf{I}+\mu^2 V_0^{2/3}F_{ab}{^o}e^a_i{^o}e^b_j,
                        \label{reg2}
\end{eqnarray}
where the fiducial triad ${^o}e^a_i$  dual to the fiducial cotriad
${^o}\omega_a^i$ defined as
$q_{ab}{^o}=\delta_{ij}{^o}\omega_a^i{^o}\omega_b^i$. Here $\mu
V^{1/3}_0$ is the coordinate length of the path along which the
elementary holonomy $h_i$ is calculated. The $\mu$ is a
dimensionless parameter which controls the length. In the limit
$\mu \rightarrow 0$, Eqs. (\ref{reg1}) and  (\ref{reg2}) become
equalities. Combining Eq. (\ref{reg1}) and Eq. (\ref{reg2}) one
can write
\begin{eqnarray}
\epsilon^{ijk} \text{tr} \left[h_{\Box_{ij}} h_k \{h^{-1}_k ,v\}  \right]  =
\nonumber  \\
-\mu^3 V_{0} \epsilon^{ijk} {^o}e^a_i{^o}e^b_j{^o}e^c_k  \text{tr}\left[F_{ab}
\{A_c, v \} \right].
\end{eqnarray}
Based on this relation with $ \epsilon^{ijk} {^o}e^a_i{^o}e^b_j{^o}e^c_k =
\epsilon^{abc}$ and
restricting the spatial integration to the fiducial volume $V_0$,
one can regularize Eq. (\ref{ApAHg}) into the form
\begin{eqnarray}
H_g^{(\lambda)} = - \frac{vN}{32\pi^2 G^2\gamma^3\lambda^3} \sum_{ijk}
\epsilon^{ijk} \text{tr} \left[h_{\Box_{ij}} h_k \{h^{-1}_k ,v\} \right],
\nonumber \\
 \label{RegHam}
\end{eqnarray}
where $\lambda^3=\mu^3 v=$ const. This condition means that the
physical size of the link $\lambda$ remains constant during the
evolution.  The choice is motivated by the correspondence  with
the classical cosmology for the large values of volume $v$ and is
equivalent to the so-called improved scheme of LQC
\cite{Ashtekar:2006wn}.  The classical unmodified Hamiltonian of
the FRW model can be recovered from $\lim_{\lambda \rightarrow 0}
H_g^{(\lambda)} =H_g$.

Inserting the elementary holonomy
\begin{equation}
h_k = \cos\left(\frac{\lambda
\beta}{2}\right)\mathbb{I}+2\sin\left( \frac{\lambda
\beta}{2}\right) \tau_k \label{hk1}
\end{equation}
and its inversion
\begin{equation}
(h_k)^{-1} = \cos\left(\frac{\lambda
\beta}{2}\right)\mathbb{I}-2\sin\left( \frac{\lambda
\beta}{2}\right) \tau_k \label{hk2}
\end{equation}
into the Hamiltonian [Eq. (\ref{RegHam})]. Next, we find that
\begin{equation}
h_k \left\{(h_k)^{-1},v\right\} =h_k 4\pi G
\frac{\partial}{\partial \beta} (h_k)^{-1} =-4\pi G \gamma \lambda
\tau_k \label{A3}.
\end{equation}
To get this relation we have used the definition of the Poisson
bracket, Eq. (\ref{Poisson1}), and  the equality
$\tau_{k}^2=-\frac{1}{4}\mathbb{I}$. Then, making use of Eq.
(\ref{A3}) turns the Hamiltonian, Eq. (\ref{RegHam}), into
\begin{eqnarray}
H^{(\lambda)}_g = \frac{vN}{8\pi G\gamma^2\lambda^2} \sum_{ijk}
\epsilon^{ijk} \text{tr} \left[h_{\Box_{ij}}\tau_k \right]. \label{A4}
\end{eqnarray}
At this stage, the relation
\begin{equation}
\text{tr}\left[h_{\Box_{ij}} \tau_k\right] = -\frac{\epsilon_{ijk}}{2}
\sin^2(\lambda\beta) \label{A5}
\end{equation}
can be applied. To derive Eq. (\ref{A5}), we insert Eq. (\ref{hk1}) and
Eq. (\ref{hk2}) into the definition of $h_{\Box_{ij}}$, and use the
formulae
\begin{eqnarray}
\text{tr}[\mathbb{I}] &=&2, \\
\text{tr}[\tau_i] &=&0, \\
\text{tr}[\tau_i\tau_j] &=&-\frac{1}{2} \delta_{ij}, \\
\text{tr}[\tau_i\tau_j\tau_{k}] &=& -\frac{1}{4}\epsilon_{ijk},
\end{eqnarray}
as well as $[\tau_i,\tau_j]=\epsilon_{ijk} \tau_k$. Afterwards, we
use Eq. (\ref{A5}) to get
\begin{eqnarray}
\sum_{ijk} \epsilon^{ijk} \text{tr}\left[h_{\Box_{ij}} \tau_k\right] &=&
-\frac{1}{2} \sin^2(\lambda\beta) \sum_{ijk} \epsilon^{ijk}\epsilon_{ijk}
\nonumber\\
 &=&-3\sin^2(\lambda\beta).\label{final}
\end{eqnarray}
Finally, inserting Eq. (\ref{final}) into Eq. (\ref{A4}) gives
\begin{eqnarray}
H^{(\lambda)}_g = -\frac{3N}{8\pi G\gamma^2}
\frac{\sin^2(\lambda\beta) } {\lambda^2}\, v. \label{Hfinal}
\end{eqnarray}

The standard Friedmann equation can be obtained from Eq.
(\ref{Hfinal}) in the limit $\beta \rightarrow 0$ (limit of small
Hubble factor) if we complete it by the matter Hamiltonian.

\section{Symplectic form} \label{Appendix2}

The symplectic form corresponding to the Poisson bracket, Eq.
(\ref{Poisson1}),  reads
\begin{equation}
\omega = \frac{1}{4\pi G \gamma} d\beta \wedge dv +d\phi \wedge dp_{\phi}.
\end{equation}
Let us find the symplectic form $\Omega$ on the surface of
constraints, namely
\begin{equation}
\Omega := \left. \omega \right|_{H^{(\lambda)}=0}.
\end{equation}
Differentiating the Hamiltonian constraint we find
\begin{equation}
-\frac{6}{8\pi G \gamma^2 \lambda} \sin(\lambda \beta)\cos(\lambda\beta)
d\beta +\frac{p_{\phi}dp_{\phi}}{v^2}-\frac{p^2_{\phi}}{v^3}dv =0.
\end{equation}
Based on this expression we obtain
\begin{equation}
\omega = \left(\frac{1}{4\pi G \gamma} \frac{v}{p_{\phi}}d\beta +d\phi \right)
\wedge dp_{\phi}. \label{B4}
\end{equation}
Differentiating  $\mathcal{O}_2$  we find
\begin{equation}
d\mathcal{O}_2 =d\phi+\frac{\text{sgn}(p_{\phi})}{\sqrt{12\pi G}}
\frac{\lambda d\beta}{\sqrt{\sin^2(\lambda \beta)-\delta}}.
\end{equation}
This expression, due to the Hamiltonian constraint, reads
\begin{equation}
\label{BB}
d\mathcal{O}_2 =d\phi+\frac{1}{4\pi G \gamma} \frac{v}{p_{\phi}}d\beta.
\end{equation}
Inserting Eq. (\ref{BB}) to Eq. (\ref{B4}) we finally get
\begin{equation}
\Omega = d\mathcal{O}_2 \wedge d\mathcal{O}_1.
\end{equation}
Thus, the Poisson bracket on $\mathcal{F}_{\text{phys}}$ takes the
form
\begin{equation}
 \left\{ \cdot , \cdot \right\} :=
\frac{\partial \cdot}{\partial \mathcal{O}_2} \frac{\partial \cdot }
{\partial \mathcal{O}_1} -
\frac{\partial \cdot }{\partial \mathcal{O}_1} \frac{\partial \cdot }
{\partial \mathcal{O}_2}.
\end{equation}

\end{document}